\documentclass[12pt]{article}
\begin{document}
\input{epsf}
\Large
\begin{center}
\bf{
Formation lengths of hadrons in lepto-production
}
\end{center}

\normalsize
\begin{center}
L.~Grigoryan
\end{center}
\begin{center}
Alikhanyan National Science Laboratory
(Yerevan Physics Institute), Br.Alikhanian 2, 375036 Yerevan, Armenia 
\end{center}
\begin {abstract}
\hspace*{1em}
The average formation lengths of the hadrons produced during the
deep inelastic scattering (DIS) of leptons on protons are studied
in the framework of the symmetric Lund model. It is shown that these
formation lengths essentially depend on the electric charges of the
hadron. For electro-production and charged current (CC)
neutrino-production, the average formation lengths of positively
charged particles are larger than those of negatively charged
antiparticles. This situation is reversed for CC
antineutrino-production. In all the mentioned cases, the main
mechanism is the direct production of hadrons. The additional
mechanism of hadron production, through the decay of resonances, is
essential only for pions and leads to a decrease in the average
formation lengths.
\end {abstract}
\twocolumn
\section{Introduction}
\normalsize
\hspace*{1em} 
The study of the space-time evolution of hadronization during
lepto-production on nuclear targets is relatively straightforward
because in this case, only one string (two jets of hadrons)
is produced. The possible influence of the two string mechanism was
studied in Ref.~\cite{akopov1}, in which its contribution was
determined to be small. Another important issue is the selection of
conditions under which cascading processes in a nucleus do not occur,
i.e., the hadronization occurs beyond the nucleus. Consequently, for
the investigation of the hadronization process, it is very important
to know the formation lengths of hadrons in lepto-production on the
elementary nuclear target (nucleon).\\  
\hspace*{1em}
Such investigations were performed in Refs.~\cite{chmaj,Bi_Gyul}.
In~\cite{Bi_Gyul}, it was claimed that for hadrons as composite
systems, the notion of formation length is ambiguous as the different
constituents of the hadrons originate at different distances.
Thus, the question of which of the following two length scales plays
the more important role in the hadronization process is an open and
model-dependent question: (i) the constituent formation length $l_c$,
which is the distance between the DIS and the first constituent
production points, or (ii) the yo-yo formation length $l_y$, which is
the distance between the DIS and yo-yo formation points, where yo-yo
is the object with the quantum numbers of the final hadron but
without its "sea".\\
\hspace*{1em}
In Refs.~\cite{akopov2,akopov3,grig3}, the formation lengths of pions
were presented in the form $l=(1 - w)l_{c}+wl_{y}$, where $w$ is the
probability of the formation length being  $l_{y}$. When $l_{c}$ and
$l_{y}$ were obtained in the framework of the Lund model, comparison
with experimental data gave $w=0.068-0.088$. This result has
confirmed the conclusion of Ref.~\cite{Bi_Gyul} on the importance of
the constituent formation length. We will further consider $l_{c}$ as
a formation length. The parameter $l_{c}$ is a function of the
variables $\nu$ and $z$ (the energy of a virtual photon and the
fraction of this energy carried away by the final hadron with energy
$E_h$ ($z=E_h/\nu$), respectively).\\
\hspace*{1em}
The aim of this work to obtain the average formation lengths of
hadrons in CC neutrino- and CC antineutrino-production on a proton
target and to compare these values with those for electro-production,
which were obtained earlier.\\
\hspace*{1em}
The paper is organized as follows. In Section 2, we briefly present a
description of the model. The results and discussion, as well as
necessary information for the calculations, are presented in Section 3.
The conclusions are presented in Section 4.
\section{Description of model}
\normalsize
\subsection{Distribution functions}
\hspace*{1em}
We begin by considering the distribution of the constituent formation
lengths $l$ of hadrons carrying away fractional energy $z$:
\begin{eqnarray}
\nonumber
D_{c}^{h}(L,z,l) = \Big(C_{p1}^{h}f(z)\delta(l-L+zL)+
\end{eqnarray}  
\begin{eqnarray}
C_{p2}^{h}\sum_{i=2}^{n}D_{ci}^{h}(L,z,l)\Big)\theta(l)\theta(L-zL-l)
\hspace{0.3cm}  ,
\end{eqnarray}
where $L = \nu/\kappa$ is the full hadronization length and $\kappa$
is the string tension.\\
\hspace*{1em}
$f(z)$ is the scaling function. It is defined by the condition that
$f(z)dz$ is the probability that the first hierarchy (rank 1) primary
hadron carries away the fraction of energy $z$ of the initial string
in the small interval $dz$. We use the symmetric Lund scaling
function~\cite{lund,seostrand}
\begin{eqnarray}
f(z) = Nz^{-1}(1 - z)^{a}exp(-bm_{\perp}^{2}/z) ,
\end{eqnarray}
where $a$ and $b$ are parameters of the model,
$m_{\perp}=\sqrt{m_{h}^{2}+p_{\perp}^{2}}$ is the transverse mass of
the emitted hadron and $N$ is a normalization factor.\\
\hspace*{1em}
Information about the functions $C_{p1}^{h}$ and $C_{p2}^{h}$ is
presented in the next subsection.\\
\hspace*{1em}
The functions $D_{ci}^{h}(L,z,l)$ are distributions of the constituent
formation lengths $l$ of the rank $i$ hadrons carrying fractional
energy $z$. To calculate the distribution functions, we use the
recursion equation of Ref.~\cite{Bi_Gyul}.
\subsection{Functions $C_{pi}^{h}$}
\hspace*{1em}
The functions $C_{p1}^{h}$ and $C_{p2}^{h}$ are the probabilities of
obtaining in some process the compositions of valence quarks for
leading (rank 1) and subleading (rank 2) hadrons on the proton target.
For hadrons of rank greater than $i=2$, the condition
$C_{pi}^{h} \equiv C_{p2}^{h}$ is fulfilled. We consider the three
types of processes: electro-production, CC neutrino- and CC
antineutrino-production. $C_{p2}^{h}$ does not depend on the process
type. The baryons in our scheme are constructed of the valence quark
and diquark. The function $C_{p1}^{h}$ is a composition of three
factors: (i) the probability that as a result of DIS, the first
constituent (quark, antiquark) of the final meson is knocked out;
(ii) the probability that as a result of the first break of the string,
the second constituent of the final meson arises; and (iii) the
probability that these partons transform into the desired meson. The
function $C_{p2}^{h}$ is composed of two factors: (i) the probability
that as a result of two consequent breaks of the string, the first and
second constituents of the final meson arise and (ii) the probability
that these partons transform into the desired meson. For final baryons,
the antiquark should be replaced by a diquark. These functions for
electro-production were presented in Refs.~\cite{grig1,grig2}, and for
(anti)neutrino-production, they can be obtained using
Refs.~\cite{grig1,grig2,barone,yao}.
\subsection{Average formation length}
\hspace*{1em}
We present here a realistic approach for the calculation of the
average formation lengths considering the type of process in which the
hadron was produced, the types of hadron and the target. Unfortunately,
for the symmetric Lund scaling function, the analytic summation of the
produced hadrons sequence over all ranks is impossible. Therefore, we
restrict ourselves to $n=10$ in eq.(1).\\
\hspace*{1em}
In the general case, the function $L_{c}^{h}$, $L_{c}^{h} = <l_{c}>$
can be written in the form:
\begin{eqnarray}
\nonumber
L_{c}^{h} = \int_0^{\infty} ldl\Big(\alpha_{h}D_c^{h}(L,z,l)
\end{eqnarray}
\begin{eqnarray}
\nonumber
+\alpha_{R}\sum_{R}D_c^{R/h}(L,z,l)\Big)/\int_0^{\infty}
dl\Big(\alpha_{h}D_c^{h}(L,z,l)
\end{eqnarray}
\begin{eqnarray}
+\alpha_{R}\sum_{R}D_c^{R/h}(L,z,l)\Big) \hspace{0.15cm}.
\end{eqnarray}
When combined with direct production, a few resonances may also
contribute. Here, $\alpha_{h}$ ($\alpha_{R}$) is the probability that
the composite parton system transforms into a hadron (resonance). We
use the condition $\alpha_{h} = \alpha_{R} = \frac{1}{2}$.\\
\hspace*{1em}
The distribution functions $D_c^{h}(L,z,l)$ and $D_c^{R/h}(L,z,l)$
were described in detail in Refs.~\cite{grig1,grig2}.
\section{Results and Discussion}
\normalsize
\hspace*{1em}
All the calculations were performed at fixed values of $E$, $\nu$ and
$Q^{2}$ equal to $27.5 GeV$, $10 GeV$ and $2.5 GeV^{2}$, respectively.\\
\hspace*{1em}
The scaling function $f(z)$ in eq.(2) has two parameters~\cite{seostrand}
$a=0.3$, $b=0.58 GeV^{-2}$ and depends on the type of produced hadron.
In the calculations, the types of observed hadrons and their parent
resonances were considered, while for other hadrons, a summation over
their types was performed. The string tension was fixed at a value of
$\kappa = 1 GeV/fm$.\\
\hspace*{1em}
The leading order parton distribution functions from~\cite{grv} were used
for the quarks (antiquarks) in the proton.\\
\hspace*{1em}
We assume that new $q\bar{q}$ pairs are $u\bar{u}$ with probability
$\gamma_{u}$, $d\bar{d}$ with probability $\gamma_{d}$ and $s\bar{s}$
with probability $\gamma_{s}$. It follows from isospin symmetry that
$\gamma_{u} = \gamma_{d} = \gamma_{q}$. For these calculations, we use
the set of values for $\gamma$~\cite{seostrand}:
$\gamma_{u} : \gamma_{d} : \gamma_{s} = 1 : 1 : 0.3$.\\
\hspace*{1em}
For baryon production, we use the set of probabilities for the production
of diquark-antidiquark pairs. For simplicity, we express them using the
probabilities of light quark-antiquark pairs production. We use:
$\gamma_{ud0}=0.1\gamma_{q}$ for the $ud$-diquark with $S=I=0$ ($S$-spin,
$I$-isospin);  $\gamma_{dd1}=\gamma_{ud1}=\gamma_{uu1}=0.015\gamma_{q}$
for light diquarks with $S=I=1$; and
$\gamma_{ds}=\gamma_{us}=0.012\gamma_{q}$ for strange diquarks.\\
\begin{figure}[!htb]
\begin{center}
\epsfxsize=8.cm
\epsfbox{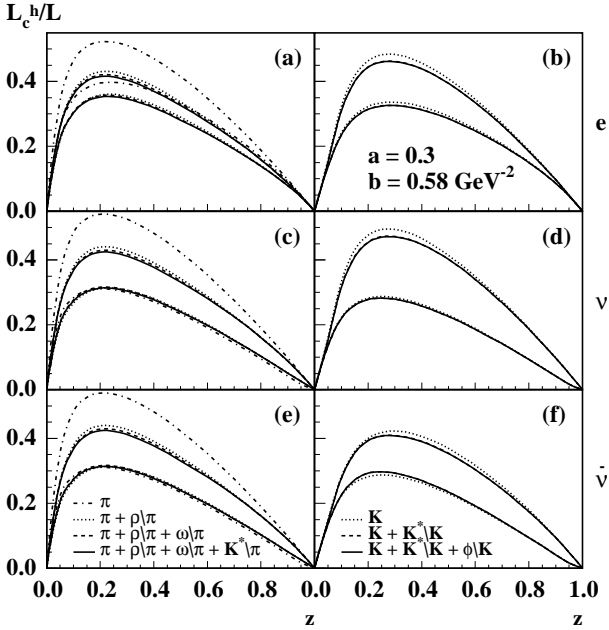}
\end{center}
\caption{\label{xx1}
{Normalized average formation lengths of pseudoscalar mesons on proton
target as functions of $z$.
}}
\end{figure}
\begin{figure}[!htb]
\begin{center}
\epsfxsize=8.cm
\epsfbox{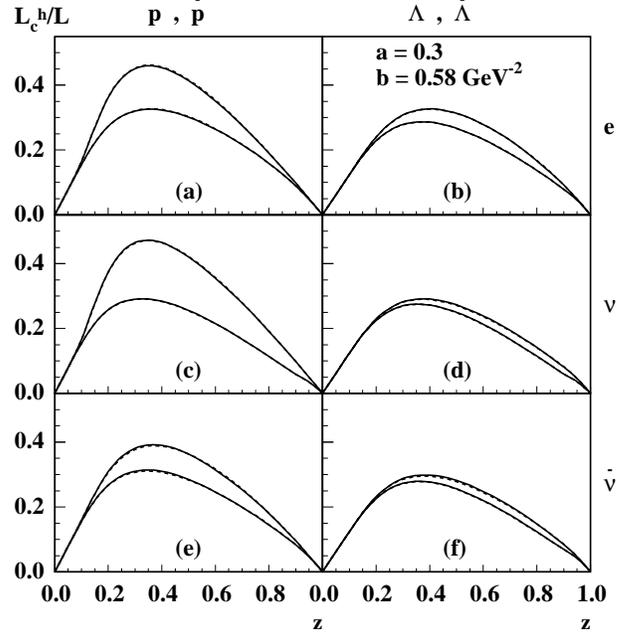}
\end{center}
\caption{\label{xx2}
{Normalized average formation lengths of baryons and antibaryons on
proton target as functions of $z$.
}}
\end{figure}
\hspace*{1em}
The normalized average formation lengths of the pseudoscalar mesons on
the proton target, calculated using the symmetric Lund model as
functions of $z$, are presented in Fig. 1 for electro- (panels a, b),
CC neutrino- (panels c, d), and CC antineutrino- (panels e, f)
production. The formation lengths are represented for pions in panels
a, c and e and for kaons in panels b, d and f. The values of the
parameters of the symmetric Lund model are also presented. The
following parameters are indicated: the formation lengths for the
direct pions (dashed-dotted lines); the sum of the direct pions and
the pions from the decay of $\rho$ mesons (dotted lines); the sum of
the direct pions and the pions from the decays of the $\rho$ and
$\omega$ mesons (dashed lines); and the sum of the direct pions and
the pions from the decays of the $\rho$, $\omega$ and $K^{*}$ mesons
(solid lines). The other indicated parameters are as follows:
the formation lengths for the direct kaons (dotted lines); the sum of
the direct kaons and the kaons from the decay of $K^{*}$ mesons (dashed
lines); and the sum of the direct kaons and the kaons from the decays
of the $K^{*}$ and $\phi$ mesons (solid lines). For
electro- and CC neutrino-production, the upper and lower curves of each
type represent the formation lengths of positively and negatively
charged hadrons, respectively. For CC antineutrino-production, the
upper and lower curves represent the formation lengths of negatively
and positively charged hadrons, respectively.\\
\hspace*{1em}
In Fig. 2, the normalized average formation lengths of baryons and
antibaryons in the symmetric Lund model as functions of $z$ are
presented in panels a and b, c and d, and e and f for electro-, CC
neutrino- and CC antineutrino-production, respectively. The results
for protons and antiprotons are presented in panels a, c and e, and
the results for $\Lambda$ and $\bar{\Lambda}$ are presented in panels
b, d and f. The contribution of  $\Delta$ ($\bar{\Delta}$) resonance
and $\Sigma$ ($\bar{\Sigma}$) resonance is considered for protons
(antiprotons) and $\Lambda$ ($\bar{\Lambda}$), respectively.\\
\hspace*{1em}
From Figs. 1 and 2, certain general features of the average formation
lengths are apparent: (i) For electro- and CC neutrino-production,
the positively charged hadrons ($\pi^{+}$, $K^{+}$ and proton) have
larger formation lengths than the negatively charged hadrons
($\pi^{-}$, $K^{-}$ and antiproton); for CC antineutrino-production,
the situation is reversed. (ii) All the curves have a characteristic
form with values equal to zero on two boundary points along $z$ and
one maximum in the vicinity of $z=0.2$ for pions, $z=0.25$ for kaons,
$z=0.35$ for protons and $z=0.4$ for $\Lambda$. The magnitudes of the
maxima of the distribution do not exceed $0.5L$. (iii) The contribution
from the decay of resonances is maximal for pions. This contribution
reaches $\sim 20\%$ for $\pi^{+}$ ($\pi^{-}$) meson for the electro-
and CC neutrino- (CC antineutrino-) production processes. As expected,
the maximal contribution originates from the $\rho$ meson. The
contribution of resonances can be neglected for kaons and baryons.
(iv) A large difference between the formation lengths of oppositely
charged hadrons is observed for kaons and protons. The maximal
difference occurs for CC neutrino-production. (v) For
electro-production, the difference between the average formation
lengths of particles and antiparticles vanishes at $z > 0.85$; for CC
neutrino- and CC antineutrino-production, the difference is large
enough for $z$ to equal unity. (vi) Unlike charged hadrons, the
formation length of a neutral baryon ($\Lambda$) is larger than that
of the corresponding antibaryon ($\bar{\Lambda}$) for all three types
of the production processes; however, the difference between the average
formation lengths of $\Lambda$ and $\bar{\Lambda}$ is very small.\\
\hspace*{1em}
Let us briefly discuss why the average formation lengths of positively
(negatively) charged hadrons are larger than those of negatively
(positively) charged ones for electro- and CC neutrino-
(CC antineutrino-) production. In electro-production, this phenomenon
occurs because of the large probability of knocking out a $u$ quark as
a result of DIS. The knocked out quark enters a composition of the
leading hadron, which has a maximal formation length. The $K^{+}$ meson
has an average formation length that is larger than the $\pi^{+}$
meson because, in the first case, the influence of resonances is
smaller. The $K^{-}$ meson has a smaller average formation length than
the $\pi^{-}$ meson because it is constructed from the "sea" quarks of
the proton and cannot be a leading hadron, whereas the $\pi^{-}$ meson
can be leading because of the $d$ quark entering its composition. In CC
neutrino- (CC antineutrino-) production, the knocked out parton obtains
a positive (negative) electric charge, which implies that the leading
hadron can be preferably positively (negatively) charged. It should be
noted that in the string model, the formation length of the leading
(rank 1) hadron $l_{c1} = (1 - z)\nu/\kappa$, does not depend on the
type of process or on the hadron or target type.
\section{Conclusions}
\normalsize
\hspace*{1em}
In this work, for the first time, the average formation lengths of
pseudoscalar mesons, baryons and antibaryons for CC neutrino- and CC
antineutrino-production on a proton target in the framework of the
symmetric Lund model have been determined.\\
\hspace*{1em}
The results of the calculations were compared with the previously
determined results of electro-production (see Refs.~\cite{grig1,grig2}).
It was shown that despite certain differences in shape and magnitude,
the average formation lengths for electro- and CC
neutrino-production have similar behaviors. CC antineutrino-production
differs from the others by the replacement of the particles and
antiparticles on antiparticles and particles, respectively.\\
\hspace*{1em}
Finally, we would like to discuss the possible application of the
obtained results. Recently, we have used the simplified version of the
average formation lengths calculation for pions obtained in the
framework of the symmetric Lund model to fit SIDIS data, in which the
possibility of daughter pions production was neglected in eq.(3) and
the condition $C_{p1}^{h}=C_{p2}^{h}=1$ was applied. These
simplifications were necessary to reduce the computation time. The
two-parametric fit yielded satisfactory agreement with the
data~\cite{grig3}.

\end{document}